\documentclass[pre,showpacs,twocolumn,nofootinbib]{revtex4}

\usepackage{graphicx}
\usepackage{epsfig}

\begin{document}

\title{A Model for Folding and Aggregation in RNA Secondary Structures}

\author{Vishwesha Guttal}
\author{Ralf Bundschuh}

\affiliation{191 W Woodruff Ave, Department of Physics, The Ohio State University, Columbus,
  Ohio 43210-1117}


\begin{abstract}
We study the statistical mechanics of RNA secondary structures designed
to have an attraction between two different types of structures as a
model system for heteropolymer aggregation. The competition between the
 branching entropy of the secondary structure and the energy gained by
 pairing drives the RNA to undergo a `{\it temperature independent}' second
 order phase transition from a molten to an {\it aggregated phase}. The
 aggregated phase thus obtained has a macroscopically large number of contacts
 between different RNAs. The partition function scaling exponent for
 this phase is $\theta \approx 1/2$ and the crossover exponent of the
 phase transition is $\nu \approx 5/3$. The relevance of these calculations to
 the aggregation of biological molecules is discussed.

\end{abstract}

\maketitle




RNA secondary structures are an excellent model system to study
folding phenomena in heteropolymers. Unlike in the protein folding
problem where a large number of different monomers needs to be taken
into account to understand folding \cite{proteinref}, an RNA has just
four bases A, U, C, and G. The interactions between these bases are
simpler than in the protein folding problem due to the separable
energy scales of the secondary and the tertiary structure. These
features make RNA secondary structures a both analytically and
numerically amenable model for rigorously studying various generic
thermodynamic properties of heteropolymer folding
~\cite{higgs00,higgs96,bund99,bund02,orland02,higgs93}.

Quite a lot is known about the folding thermodynamics of {\em
single} RNA molecules. At low temperatures, where monomer specific
binding energies and sequence heterogeneity are important, the
resulting (frozen) phase is {\it glassy} \cite{higgs96,bund02}. At
high temperatures, thermal fluctuations lead to a {\it
denatured phase}, where the backbone is randomly coiled (without any
binding) like a self-avoiding random walk. At intermediate
temperatures, where an effective attraction between short segments is
important, the molecules are expected to be in the so called {\it
molten phase} \cite{bund99,bund02}. In this molten phase
many different secondary structures all having comparable
energies (within $O(k_BT)$) coexist. If
the tendency of biological sequences to be designed to fold into a
specific, functional structure is taken into account, the {\it native
phase} emerges~\cite{higgs93,bund99}. Many important questions have
been raised with regard to these phases, e.g. their stability,
characteristics, and the properties of the phase transitions between
them in the context of both protein and RNA folding
~\cite{proteinref,higgs00,higgs96,bund99,bund02,orland02,higgs93}. In
this Letter we shall begin to understand another important aspect of
heteropolymer folding, namely the competiton between the individual
folding of the molecules and {\it aggregation} of several molecules,
using the RNA secondary structure formulation.


In the context of protein folding, the competition between individual
folding and misfolding associated with aggregation is a very important
phenomenon.  The failure of protein molecules to fold correctly and
the associated formation of alternative structures stabilized by
aggregation is associated with various diseases such as
Alzheimer's, Mad Cow, and
Parkinson's~\cite{prusiner99,dobson03}. Thus, this phenomenon has been
studied with the tools available for the protein folding problem in
various contexts~\cite{dobson03,slepoy}. But also in the realm of RNA
folding the competition between individual and aggregated
structures plays an important role, e.g., in the growing field of
riboswitches~\cite{storz05}. In these riboswitches, the aggregation of
two RNA molecules through base-pairing in competition with the base
pairing of the individual molecules is used to regulate the expression
of genes in dependence on the concentrations of the RNAs
involved. Even the local structure of double stranded DNA in the
repeat regions of the genes involved in triplet repeat deseases
(Huntingtons, fragile X, etc.~\cite{mitas97}) is an
example of an aggregated structure (the double stranded DNA) competing
with the multitude of secondary structures the single strands of this
DNA can form by themselves since their repeat units of, e.g., CAG and
CTG in Hungtington's desease, allow self-pairing as well. Here, we
approach the phenomena associated with competition between
intra-molecular structure and aggregation by considering a toy model
to study the phase transition of an RNA secondary structure from the
molten to an aggregated phase. While our model is
literally applicable to the above mentioned triplet repeat
desease genes, we see it more broadly as a basic model for studying
the competition between intra-molecular structure and aggregation into
which later aspects of the other scenarios discussed above such as
native states and simultaneous aggregation of
several molecules can be incorporated. In studying our model, our
focus is on the thermodynamic properties of the system. Thus, we solve
the model exactly in the thermodynamic limit and calculate the
critical exponents relevant to the phase transition.


RNA is a biopolymer with four different monomers A, U, C and G in its
sequence. The Watson-Crick pairs A-U and C-G are energetically the
most favorable pairs while G-U is marginally stable and the other
combinations are prohibited. By an RNA secondary structure, we mean a
collection of binding pairs $(i,j)$ with $1 \le i < j \le N$, where N
is the number of bases in the sequence. Any two pairs $(i_1,j_1)$ and
$(i_2,j_2)$ are either nested, i.e. $i_1 < i_2 < j_2 < j_1$ or
independent i.e. $i_1 < j_1 < i_2 < j_2$. The above restriction means
we are not allowing pseudo-knots, which are generally energetically
not as favorable \cite{tinoco99}. Such a secondary structure can be
represented by a helix diagram, non-crossing arch diagram
or a mountain representation as shown in Fig.~1.

\begin{figure} \label{fig:rna}
\includegraphics[width=1.0\columnwidth]{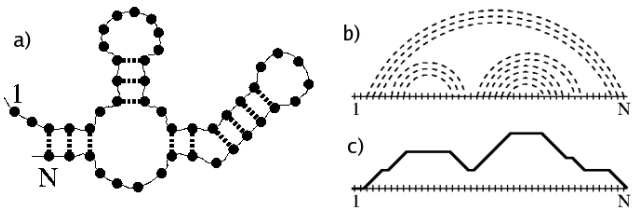}
\caption{Abstract representations of RNA secondary structures (from~\cite{bund02}). (a) Helix representation (b) Non-crossing  Arch diagram. Here, the solid line corresponds to the backbone of the RNA. The dashed arches correspond to the base pairs. The absence of pseudo-knots implies that the arches never cross. (c) Mountain Representation. Here, as we go along the backbone of the RNA from base 1 to N (represented by the base line), we go one step up for the beginning of a pair, one step down for the closing of a pair and a horizontal step for no pairing. Such a mountain never crosses the baseline and always returns to the baseline at the end.}
\end{figure}

Let the free energy associated with the pairing of bases $i$ and $j$ in an RNA be $\epsilon_{ij}$. This free energy has contributions from the gain in the energy due to binding and the associated configurational entropy loss. In addition to these, in principle there are also entropic and/or energetic effects due to loop formation, stacking, etc. Even though the accurate parameters as determined by the experiments are essential to calculate the exact secondary structure, such microscopic details as well as the exact values of the energies $\epsilon_{ij}$ do not affect the asymptotic properties of the phases and the critical exponents. Hence, we ignore them in our model calculations.


To understand the phase transition from the molten to the aggregated
phase, we first define the aggregated phase as an ensemble of RNA
secondary structures in which a macroscopically large number of
contacts occur between two different RNAs. We consider a dual RNA
biomolecule system consisting of two types of RNA in a solution. We
refer to them as RNA-1 and RNA-2. Individually, RNA-1 and RNA-2 are in
the molten phase. However, when they are together in a solution, also
base pairings between bases from different molecules are possible.
We study the phases of this dual RNA system, as the bias
strength is varied.

To do so, we assume a simple pairing energy model with the free energy of pairing between bases $i$ and $j$ defined as:
\begin{equation} \label{eq:model}
\epsilon_{i,j} = \left\{ \begin{array}{ll}
    \epsilon_1 & \textrm{if } i,j \in \textrm{RNA-1} \\
    \epsilon_2 & \textrm{if } i,j \in  \textrm{RNA-2} \\
    \epsilon_3 & \textrm{if } i \in \textrm{RNA-1,}  j \in  \textrm{RNA-2  or vice-versa}   \\
\end{array} \right.
\end{equation}
Here, the intra RNA base pairing energies $\epsilon_1$ and
$\epsilon_2$ could be of comparable magnitude in a realistic RNA
molecule. The inter RNA base pairing energy, or the bias, $\epsilon_3$
is the parameter which can in principle be controlled by sequence
mutation. Note, that neglecting sequence heterogeneity in this kind of
models was established as a useful approximation at not too low
temperatures in a similar context~\cite{bund99} (see also the
discussion at the end of this letter).

Denote the Boltzmann factors corresponding to the pairing energies by $q_1$, $q_2$ and $q_3$ respectively. We show that this simple model predicts a molten to an aggregated phase transition, as we tune the parameter $q_3$. We do
so by exploiting the recursive relation ~\cite{mccaskill90,water78}
\begin{equation} \label{eq:rec}
Z_{ij} = Z_{i,j-1} + \sum_{k=i}^{j-1} Z_{i,k-1} e^{-\epsilon_{jk}/T} Z_{k+1,j-1}
\end{equation}
for the partition function $Z_{ij}$ for a sequence of bases from $i$ to $j$,
which can be evaluated in $O(N^3)$ time starting from the
initial conditions $Z_{i,i}=Z_{i,i-1}=1$.


 To keep the analytical calculations simple, we assume each RNA to be of equal length, containing $N-1$ bases \cite{equallength}. We now consider the joint folding of these two RNAs and denote its partition function by $Z_d(N;q_1,q_2,q_3)$. As explained before, the free energy of pairing for the bases belonging to a given RNA has contributions from the energy gain due to the pairing and the entropy loss associated with the loop formation. This holds true even for pairing across the bases belonging to different RNAs. But when the first pairing between the bases belonging to different RNAs occur, there is an additional entropic loss due to the breakdown of translational invariance symmetry. Thereafter, only the free energy $\epsilon_3$ plays a role in the inter RNA base pairing. In the thermodynamic limit, this additional entropic loss has no effect on the phase of the system, but it is the energetics of pairing that drives the phase transition. Hence, we ignore this additional entropic term. This essentially reduces the problem to the folding of a single sequence with $2N-2$ bases. The aggregated secondary structure can now be interpreted as having a macroscopically large number of contacts between the two halves of the concatenated RNA.


Let us first consider two special cases. Setting $q=q_1=q_2=q_3$ corresponds to the well known molten phase of the RNA secondary structure, whose partition function can be calculated exactly in the asymptotic form $Z_d(N;q,q,q) = Z_0(2N;q) = A(q)(2N)^{-\theta}z_c(q)^{2N}$
with the characteristic scaling exponent $\theta=3/2$ \cite{bund99}. This exponent is characteristic in the sense that it is insensitive to various microscopic details of the RNA secondary structure such as the cost of a hairpin loop, weak sequence heterogeneity, etc. The other simple case is $q_3=0$. This case describes two RNAs in the molten phase which do not know of each other's presence. The partition function of such a dual RNA is then just the product of individual partition functions, i.e. $Z_d(N;q_1,q_2,0) \equiv Z_0(N,q_1)Z_0(N,q_2)$. Hence the scaling exponent is $\theta=3$.

We now want to understand the case of general $q_1$, $q_2$ and $q_3$. To this end we calculate the partition function of the dual RNA as follows. Let the base pairings within a given RNA be called primary and those across different RNAs be called secondary. Any given secondary structure thus obtained has a series of secondary pairings ($i_1$,$j_1$),\ldots,($i_k$,$j_k$) such that $1 \leq i_1 <\ldots < i_k \leq N-1$ and $1 \leq j_1 <\ldots< j_k \leq N-1$. Note that we have labeled the RNA-1 by $i$ and the RNA-2 by $j$ indices. The bubbles thus formed between any two consecutive secondary pairings are allowed to have only the primary pairings. If all the secondary structure configurations are enumerated according to the number of the inter-RNA (or the secondary) contacts $k$, then the total partition function of this dual RNA system, in the $z$-transform representation can be written as:
\begin{eqnarray} \label{eq:partition}
\hat{Z_d}(z; q_1,q_2,q_3) = \sum_{k=0}^\infty q_3^k \hat{Z_0}(z;q_1)^{k+1}*\hat{Z_0}(z;q_2)^{k+1}\\
                         = \oint \frac{dz'}{z'} \frac{\hat{Z}_0(z';q_1)\hat{Z}_0(z/z';q_2)}{1-q_3\hat{Z}_0(z';q_1)\hat{Z}_0(z/z';q_2)}
\end{eqnarray}
where $\hat{Z}_d(z;q_1,q_2,q_3)$ and $\hat{Z}_0(z;q)$ are the $z$-transforms of $Z_d(N;q_1,q_2,q_3)$ and $Z_0(N;q)$ respectively. The symbol $*$ indicates the convolution in $z$-space defined as $f*g = \oint \frac{dz'}{z'} f(z')g(z/z')$. Eq.(4) is obtained by summing up the geometric series in Eq.(3). The convolution integration can be done numerically to obtain the singularities of $\hat{Z_d}$ and hence, the asymptotic behavior of $Z_d(N;q_1,q_2,q_3)$.


\begin{figure} \label{fig:results}
\includegraphics[width=0.9\columnwidth]{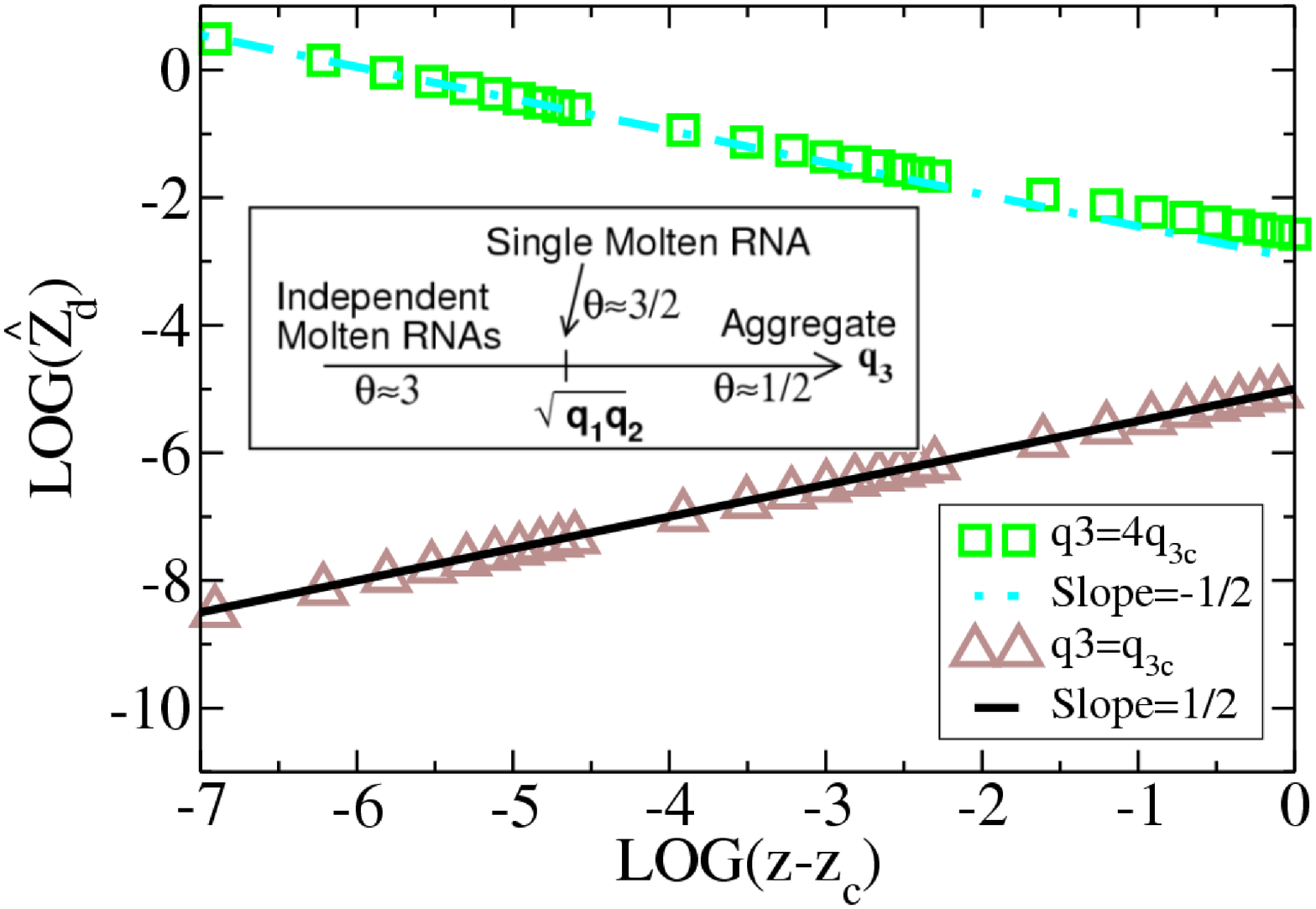}
\caption{(color online) The behavior of the partition function $\hat{Z}_d(z;q_1=4,q_2=9,q_3)$. For $q_3=q_{3c}=6$, we observe a square root behavior. For $q_3>q_{3c}$, we see an inverse square root behavior. The inset shows the resulting phase diagram.}
\end{figure}

The results are shown in Fig.~2. For $q_3=q_{3c}=\sqrt{q_1q_2}$, we find a square root singularity and hence $\theta=3/2$ \cite{singularity}, the characteristic exponent of the molten phase. For $q_3>q_{3c}$, $\hat{Z}_d$ has an inverse square root singularity, indicating a new phase. We interpret the new phase with the partition function scaling exponent $\theta \approx 1/2$ as the aggregated phase.
We claim that for all $q_3 < q_{3c}$, the dual RNA system is just the phase corresponding to $q_3=0$ in the asymptotic limit, hence $\theta=3$. This claim is verified by numerical calculations of the exact partition function for finite length and the calculation of an asymptotic macroscopic quantity (the order parameter) to be defined below. The resulting simple phase diagram is shown in the inset of Fig.~2.


In order to verify that the phase transition indeed happens at
$q_{3c}=\sqrt{q_1q_2}$, we calculate the order parameter of the
phase transition. Here, the order parameter $Q$ is defined as the
fraction of secondary pairings in a secondary structure, an
important structural property of the aggregate. For arbitrary $q_3$
the order parameter can be calculated exactly from $Q=- \lim_{N
\rightarrow \infty} (d\ln Z_d/d\ln q_3)/N$. The inset of Fig.~3
clearly shows $Q=0$ for $q_3 \leq q_{3c}$ and continuously
increasing with $q_3$ thereafter saturating to $Q=1$ for
$q_3/q_{3c}\gg 1$. From this behavior of the order parameter we can
conclude that the phase transition indeed occurs at
$q_{3c}=\sqrt{q_1q_2}$ and that the phase transition is of second
order. Physically, we can understand the behavior of the order
parameter by using the mountain representation of RNA (see Fig.~1c).
 Between any two consecutive secondary pairings, the contribution of primary pairs to the height of the mountain
  is zero. Hence, the total number of secondary pairings is equal to the height $\langle h \rangle$ of the mountain at its midpoint. Using the random walk analogy \cite{bund02,feller}, we find that $\langle h \rangle \sim O(N^{1/2})$, hence $Q \sim O(N^{-1/2})$. For $q_3 < q_{3c}$, the secondary pairings are even less likely, and hence in the thermodynamic limit $Q=0$ for $q_3 \leq q_{3c}$, consistent with what we have obtained by exact expression.



 To further verify our claims about the phase for $q_3<q_{3c}$ and to
 calculate the scaling exponents corresponding to the second order
 phase transition, we iterated the recursion relation (Eq.(1)) to
 calculate the exact partition function for RNA of finite length $N$.
 The results of the numerical calculations are in complete agreement
 with the phase diagram of Fig.~2 (inset) when extrapolated to the thermodynamic
 limit, thus verifying our claim ~\cite{sequence}. Next we calculate the free energy per
 length $f(q_1,q_2,q_3)=-\ln Z_d(N)/N$, taking into account the finite
 size effects. We assume the usual scaling function for the order
 parameter $Q(N)=N^{-1/2}g[(q_3-q_{3c})N^{1/\nu}]$ close to the critical
 point. Fig.~3 shows the result of scaling plot, with the best fit value
 for the crossover critical exponent $\nu \approx 5/3$.

\begin{figure} \label{fig:scaling}
\includegraphics[width=0.8\columnwidth]{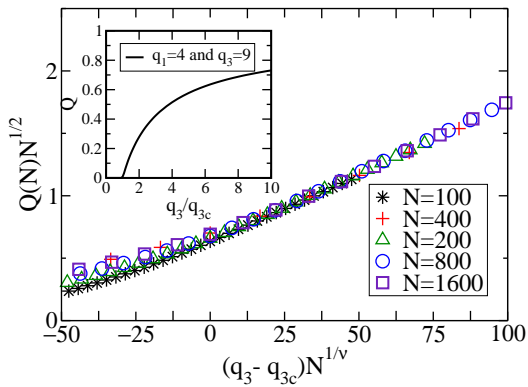}
\caption{(color online) Scaling plot for the order parameter. Inset shows the order parameter of the phase transition. In both the plots $q_1=4 \textrm{ and } q_2=9$, hence $q_{3c}=6$. }
\end{figure}




This model has some similarities with G\=o-like model studied by Bundschuh and Hwa \cite{bund99} which shows
a molten-native transition. The physics behind the phase transition in their model as well as our model is
the same, i.e., the competition between the energetic gain of the secondary contacts (or native contacts of
G\=o-like model) and the branching entropy. But, contrary to the native phase where the ground state is unique,
 the aggregated phase has degenerate ground states. On the other hand, both these models can `melt' from
 their (aggregated or native) ground state to any of the molten, glassy or denatured phase, depending
 on the temperature and the strength of the bias. The differences in the behavior of these models arises
 from the fact that for the G\=o-like model the bias is site specific where as for the model we have
 presented, the bias is towards a macroscopically large number of sites.

We would like to emphasize two simplistic considerations
of our model. Firstly, choosing a molten phase to start with, in which
the sequence heterogeneity is unimportant. It is important to note
that the bases as we call them here are not necessarily single bases,
but short segments of the sequence (such as CAG) whose effective
interaction with any other segment is the same \cite{bund99}. Even if
we do have weak sequence heterogeneity, we do not expect that such
microscopic details alter the thermodynamic results
presented here based on previous
work~\cite{bund99,bund02,sequence}. However, at low enough
temperatures where such a homogeneous approximation is no longer
valid, it should be interesting to consider the role of a suitably
defined bias in the glassy phase.  Secondly, we considered only two
RNA molecules for aggregation. In the case of multiple RNAs participating
in the folding, the ground state would
depend on how the different types of RNAs are aligned to fold. In
fact, these ground states could be topologically different from the
ground state of our two RNA model. Hence, the values of the critical
exponents for the transition might change, though the qualitative
physics of aggregation, such as the critical inter-molecular base
pairing energy at which the transition takes place would remain the
same.

In summary, we have presented a simple model for heteropolymer
folding using the RNA secondary structure formulation, which shows a
second order phase transition from an {\it independently molten} to
an {\it aggregated phase}.
 The behavior at criticality turns out to be the molten phase for the concatenated molecule. The
 transition is completely driven by the energetics of pairing and is temperature independent.
  Proteins are known to undergo a folding transition
 from the native to an aggregated phase instead of from a molten to an aggregated phase ~\cite{dobson03}. It should
be interesting to see if this study can be extended to understand
the thermodynamics of such a phase transition. It should also be
interesting to study the role of kinetics of RNA folding in this
phase transition.

We gratefully acknowledge useful discussions with Tsunglin Liu. RB is supported by the National Science Foundation through grant no. DMR-0404615.



\begin{thebibliography}{99}
\bibitem{equallength} The equal length approximation is not necessary. In general as long as the RNAs are of the order of same length, all the subsequent results hold.
\end{thebibliography}
\end{document}